# Kerr Superoscillator Model for Microresonator Frequency Combs

Jonathan M. Silver[1], Changlei Guo[1,2], Leonardo Del Bino[1] and Pascal Del'Haye[1*]

[1]*National Physical Laboratory (NPL), Teddington, TW11 0LW, United Kingdom*
[2]*School of Information Science and Engineering, Xiamen University, Xiamen 361005, China*
[*]*Corresponding author: pascal.delhaye@npl.co.uk*

Microresonator-based optical frequency combs, or "microcombs", have attracted lots of attention in the last few years thanks to their promising applications in telecommunications, spectroscopy and optical clocks. The process of comb generation in microresonators can be modelled in the frequency domain using coupled mode equations, and has also recently been successfully described in the time domain using a nonlinear Schrödinger equation known as the Lugiato-Lefever equation. Though time-domain approaches have brought many interesting insights for the understanding of microcombs, an intuitive frequency-domain model has not yet been established. In this work we present a frequency-domain model of microcombs that describes the overall structure of the spectra in terms of a few collective excitations of groups of neighboring comb lines, which we term "superoscillators". This approach ties in nicely with the recently-developed time-domain model based on soliton crystals, and links the microcomb generation process with frequency response theory.

Subject Areas: Optics, Nonlinear Dynamics

## I. INTRODUCTION

The process of optical frequency comb generation in microresonators [1], [2] can be modelled in the frequency domain using coupled mode equations [3], [4]. There have been several successful efforts in recent years to describe frequency comb generation in nonlinear optical microresonators in the time domain using a nonlinear Schrödinger equation known as the Lugiato-Lefever equation [5–9]. The dynamic balance between the driving force and dissipation on the one hand, and the nonlinearity and dispersion on the other, gives rise to short, intense pulses of light known as dissipative solitons [7], [10–14]. Time-domain approaches have brought many interesting insights for the understanding of microresonator-based frequency combs ("microcombs") [15]. Apart from the time domain models, an intuitive frequency domain model might help to get a quicker and more intuitive understanding of microcomb states using frequency domain data e.g. from an optical spectrum analyzer. Similar to the shift from mode locked lasers to frequency combs, this might provide helpful insights for the physics of microcomb generators.

In this work we introduce a frequency-domain model that gives an intuitive picture of microcombs. We demonstrate how the presence of intense circulating light gives rise to a parametric modulation of the resonance frequencies of the optical modes, which in turn influences the resonant coupling condition of the optical pump field and through this the circulating pattern of light itself. This leads to a very specific modulation pattern of the microresonator's resonance frequency. The link to the time domain can be seen as the resonance frequency periodically "chopping" apart the continuous wave pump laser to generate a soliton pattern. Such periodic soliton patterns have recently been described [16]. The viewpoint in which an "oscillating resonance frequency converts a CW laser into pulses" and that of "circulating solitons inducing a resonance frequency modulation" resemble an interesting "chicken-and-egg" problem that might shine new light on the underlying science of microcombs. Our new model can describe an entire microcomb spectrum containing hundreds of modes using a few superoscillators, which in fact seem to be strongly linked to each other, reducing the total number of degrees of freedom to typically just two or three. This superoscillator model brings together concepts from both electrical and mechanical engineering and applies them to nonlinear optics. It can also be applied more broadly to any system comprising generalized coupled parametric oscillators, and as such holds great promise as a starting point for modelling of nonlinear systems. The general interest in frequency domain descriptions of microcomb formation is also reflected in recent work based on the Kuramoto model [17] and a reduced phase model [18].

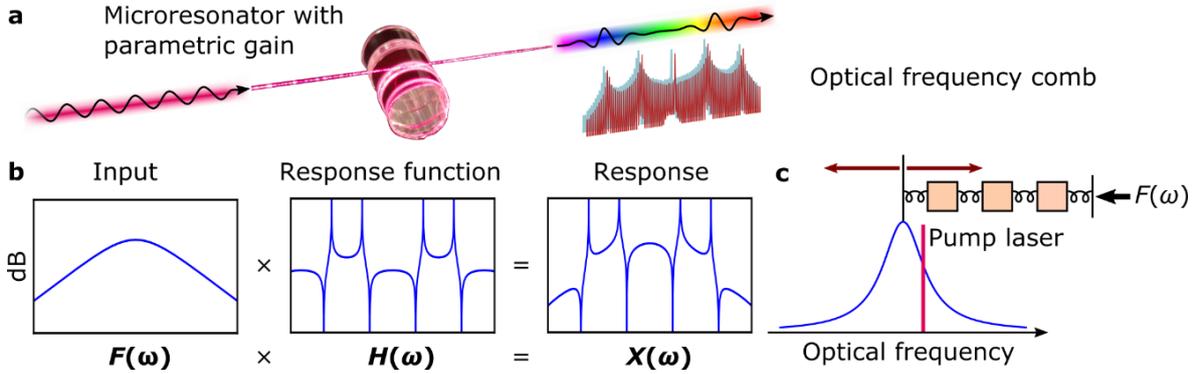

**Figure 1** Illustration of the formation of frequency combs within microresonators and the application of the frequency response function to this process. (**a**) An optical microresonator is pumped with CW light, generating a frequency comb via four-wave mixing. (**b**) The envelope of the comb spectrum can be described as the response $X(\omega)$ of the microresonator, which is given by the force term $F(\omega)$ multiplied by the frequency response function $H(\omega)$. Shown here are the magnitudes of these three functions on a dB (logarithmic) scale. $F(\omega)$ takes the form of a hyperbolic secant, which comes from the interaction of the Kerr nonlinearity with dispersion in the microresonator, and corresponds to the spectrum of a Kerr soliton [7]. $H(\omega)$ is a function containing poles and zeros ("positive and negative superoscillators"), characteristic of linear response functions of complex systems with resonances and antiresonances. (**c**) The complex waveform circulating in the resonator causes a rapid and convoluted modulation of the instantaneous optical resonance frequency via the dynamic Kerr shift, which is depicted here as $F(\omega)$ acting on the resonance frequency via a system of masses and springs. The shifting of the resonance relative to the fixed pump frequency leads to a rapid modulation of the pump power coupled into the resonator.

Frequency response functions are a tool of fundamental importance in both physics and engineering as they allow us to express the time-dependent response of a system to stimuli in the frequency domain [19]. Especially in electrical engineering and mechanical engineering these are frequently used to solve differential equations describing complex mechanical/electronic systems or in network theory. In particular, solving differential equations is significantly simplified by applying a Laplace transformation and determining complex eigenmodes instead of calculating time-domain solutions. Examples are the eigenmodes and stability regions of airplanes that are commonly described by frequency transfer functions. These concepts can be applied to microresonator-based frequency combs by starting with the response of a system to a generalized force:

$$F(\omega) \times H(\omega) = X(\omega) \qquad (1)$$

Here, $F(\omega)$, $H(\omega)$ and $X(\omega)$ are the force term, response function and response respectively. Figure 1 shows how this concept may be applied to an optical microresonator. The interplay between microresonator dispersion and parametric gain leads to a continuous spectrum $F(\omega)$ with the form of a hyperbolic secant as observed in soliton generation in microresonators. The complex parametric processes within the resonator can be described by a simple frequency response function $H(\omega)$ that generates a microcomb spectrum $X(\omega)$. Just as in a wide range of linear systems spanning from structural dynamics to electronics, in phase-locked combs $H(\omega)$ takes the form of the quotient of two polynomials:

$$H(\omega) = K \frac{(\omega - z_1)(\omega - z_2) \cdots (\omega - z_m)}{(\omega - p_1)(\omega - p_2) \cdots (\omega - p_n)} \tag{2}$$

where $z_i$ and $p_i$ are the (generally complex) positions of zeros and poles respectively. When $|H(\omega)|$ is plotted on a logarithmic scale, poles and zeros appear as inverted versions of each other, and the closer $z_i$ and $p_i$ are to the real axis, the sharper these features are. Both entail a phase shift of $\pi$ between lower and higher values of $\omega$.

In microresonators, $H(\omega)$ describes the response of the resonator to the pump light in the presence of parametric gain, and depends upon the pump power and detuning as well as the resonator's mode structure. The similarities between the observed comb spectra and linear response functions are very striking, and might hint at some physical background in which the microresonator-comb dyad behaves like a complex response system. The poles and zeros in $H(\omega)$ can be seen as positive and negative superoscillators in microcombs. These superoscillators consist of multiple comb modes and follow the envelope of poles and zeros in a response function.

Recent measurements show that the presence of solitons circulating in microresonators cause resonance frequency splittings induced by dynamic changes of the Kerr effect with the circulating pulses [20]. In more complex comb states this leads to a dynamic modulation of the resonance frequency. This effect is illustrated in Figure 1c, in which the center frequency of one of the resonator's modes is being modulated back and forth by the forcing term $F(\omega)$ after it has passed through a complex system represented here by a series of masses and springs. This shift can cause a strong amplitude modulation, or chopping, of the coupled pump light as the resonance moves relative to the laser frequency. This in turn affects the pattern of light in the resonator and thereby $F(\omega)$.

## II. MATERIALS AND METHODS

The setup we use to generate frequency combs and measure their amplitude and phase spectra is shown in Figure 2a. Light from a tunable external-cavity diode laser (ECDL) is boosted to ~100 mW by an erbium-doped fiber amplifier (EDFA) before being coupled into a silica microrod or disk resonator ($Q \approx 10^8$, free spectral range $\approx 20$ GHz) via a tapered fiber. The amplitude spectrum is measured by sending the light transmitted by the tapered fiber directly into an optical spectrum analyzer (OSA) (this step is not shown in Fig. 2a). To measure the phase spectrum (see [21], [22] for a full description of this method), the same light is instead sent through a second EDFA, followed by a liquid-crystal-based waveshaper, as shown in the schematic. The waveshaper enables individual adjustments of the amplitudes and phases of the different comb lines. Part of the light after the waveshaper is sent to an OSA, and the amplitudes of all the comb lines at this point equalized via the waveshaper. The rest of this light is sent into a home-built peak power detector consisting of a frequency-doubling crystal and a photodiode measuring the average power of the generated second harmonic. We then adjust the phases of the comb lines to maximize the signal on the peak power detector. This occurs when the adjusted phases are all equal, meaning that the optical power in the repeating waveform is compressed into a single short pulse. Once the peak power is maximized, we record the phase shifts applied by the waveshaper, which must be equal to the negative of the original comb phases.

## III. RESULTS AND DISCUSSION

The origin of superoscillators can be explained as shown in Figure 2b. The detunings of the resonator modes relative to the equidistant frequencies of a phase-locked comb follow a positive parabola due to the anomalous quadratic dispersion of the microresonator. This has been measured in the same type of microrod resonator, as shown in Fig. 5a of Ref. [22]. The symmetry of the comb about the pump frequency dictates that the detuning parabola is centered here. As a result of the Kerr-effect-induced frequency splitting of microresonator modes (linked to the complex circulating light pattern), a comb state can be have multiple detuning parabolas simultaneously. This has been recently demonstrated as a Kerr-effect-induced resonance frequency splitting as shown in Ref. [20], which corresponds to a dynamic change in the detuning. In single- and few-soliton combs, the pump is actually red-detuned most of the time [7], and only becomes blue-detuned whilst a soliton is passing due to the immense Kerr shift produced by the soliton. By contrast, in the comb states described here, the pump is blue-detuned from the resonance even in the absence of Kerr shift, meaning that the parabola always crosses zero. Previous measurements [22] have only measured the average detuning (in the backwards direction).

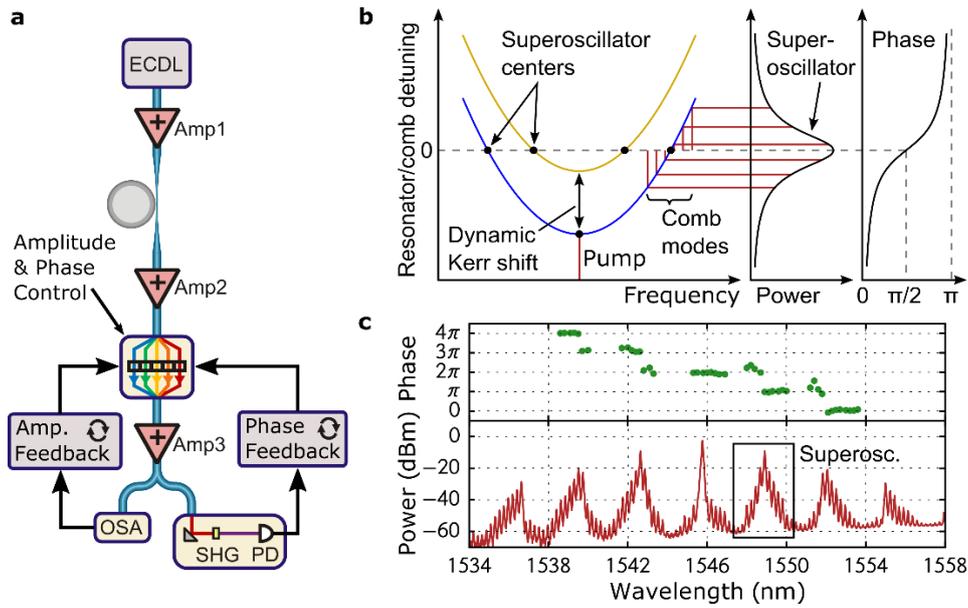

**Figure 2** (**a**) The experimental setup used to measure the combs' phase spectra (see [21], [22] for a full description of this method). (**b**) Generation of superoscillators around a zero crossing of the detuning parabola for a comb in which the pump is blue-detuned from its cavity resonance. The detuning parabola represents the detuning $\delta_{rc}$ of the various resonator modes relative to their corresponding comb modes in a resonator with pure quadratic anomalous dispersion and a pump frequency that is blue-detuned from its resonance. The parabola moves down, e.g. from the yellow to the blue curve, as a result of the instantaneous Kerr shift from a passing light pattern in the microresonator. For each parabola, superoscillators are generated around the points where the detuning crosses zero, as the parametric gain varies with $\delta_{rc}$ as the Lorentzian $1/(\delta_{rc}^2 + \gamma^2)$ where $\gamma$ is the resonator's half linewidth. (**c**) Measurement of the amplitudes and phases of the lines of a bunched comb, showing a series of superoscillators. Each superoscillator exhibits a phase shift of π between the highest- and second-highest-amplitude lines, exactly as predicted by the model.

Since this parabola specifies the detuning between any pair of sidebands equally spaced around the pump and their resonator modes, the parametric gain of the resonator is highest around its zero-crossings. Consequently this is where the first sidebands appear, which in turn mix with the pump and themselves to create a cascade of sidebands known as a primary comb [23]. In the time domain this corresponds to a modulation instability that grows into a regular array of solitons, or Turing pattern [24]. These combs are typically the first to appear as the pump laser frequency is swept down towards the resonance from above. As the pump is swept closer to the resonance, causing more power to be coupled into the microresonator, sidebands of the pump appear in several consecutive modes around these zero-crossings. The amplitudes of the sidebands in these bunches approximately follow a Lorentzian profile that is imposed by the approximately linearly changing detuning from their respective Lorentzian-shaped cavity modes. This mirrors the response function of a system in the vicinity of a resonance, or pole, motivating the use of the term "superoscillator" as a collective excitation of multiple comb modes behaving like a single harmonic oscillator. A notable feature of this simple model is that it predicts a phase shift of π between frequencies either side of the resonance, as seen on the right of Figure 2b.

Figure 2c shows a comb spectrum taken for a pump detuning just low enough for superoscillators to appear around the primary comb lines. The measured phases exhibit π phase shifts across the peaks of the resonances as predicted by the model. Whilst the superoscillators closest to the pump are centered at the zero-crossings of the detuning parabola without Kerr shift, the others can be explained by dynamic Kerr shifts. As the parabola moves down in response to a passing pulse of light, its zero crossings move to new positions further from the pump, generating additional superoscillators.

Figure 3 shows different combinations of superoscillators together with experimental data and fits of our model. These states arise from bunched comb states (Fig. 2c) when transitioning to smaller pump laser detunings. Besides positive superoscillators we observe inverted "negative" superoscillators that can be explained by destructive interference of two positive superoscillators that are π out of phase with each other. Such a negative superoscillator is a feature well known as anti-resonance in coupled oscillator systems. Our measurements are in excellent agreement with the fitted curves based on equation (2).

The positions of the various poles and zeros in our phase-locked comb states are almost always linked, forming patterns that approximately repeat after a large number of comb lines (typically of the order of 50). This may be explained by the fact that in the presence of periodic soliton patterns with localized defects [16], $H(\omega)$ can be expressed as one of the following two forms (see Appendix 1):

$$H(\omega) = K \frac{\sin(\alpha\tau\omega/2)}{\sin(\tau\omega/2)} \qquad (3)$$

$$H(\omega) = K \left( \frac{\sin(\alpha\tau\omega/2)}{\sin(\tau\omega/2)} - \exp(i\beta\tau\omega/2) \right) \qquad (4)$$

We may write (3) in the form of (2) as follows using the well-known product expansion of $\sin x$:

$$H(\omega) = K\alpha \prod_{n \neq 0} \left( \frac{\alpha \left( \omega - \frac{2n\pi}{\alpha\tau} \right)}{\omega - \frac{2n\pi}{\tau}} \right) \qquad (5)$$

A similar approach may be applied to (4). It is clear from this that the positive superoscillators are equally spaced by $2\pi/\tau$ in $\omega$ or $f_{\text{PS}} = 1/\tau$ in frequency.

The power spectra in Figure 3 were fitted with response functions of the form of either (3) or (4) multiplied by a hyperbolic secant, the results of which are given in Table 1. We present $f_{\text{PS}}$ in place of $\tau$, and also include for comparison the comb mode spacing (repetition rate) $f_{\text{rep}}$ and the ratio $f_{\text{PS}}/f_{\text{rep}}$.

The fitted curves agree well with the measured data (see Appendix 2). Recent results and simulations based on the Lugiato-Lefever equation suggest that these fits and observations could be explained by positing that the resonator is completely filled with a series of equally-spaced solitons with a few defects such as a single anomalous spacing or a shifted soliton. According to this model, $\tau$ is the spacing

(in time) between successive solitons in the regular lattice, and so $f_{PS}/f_{rep}$ is the ratio between the round-trip time and this.

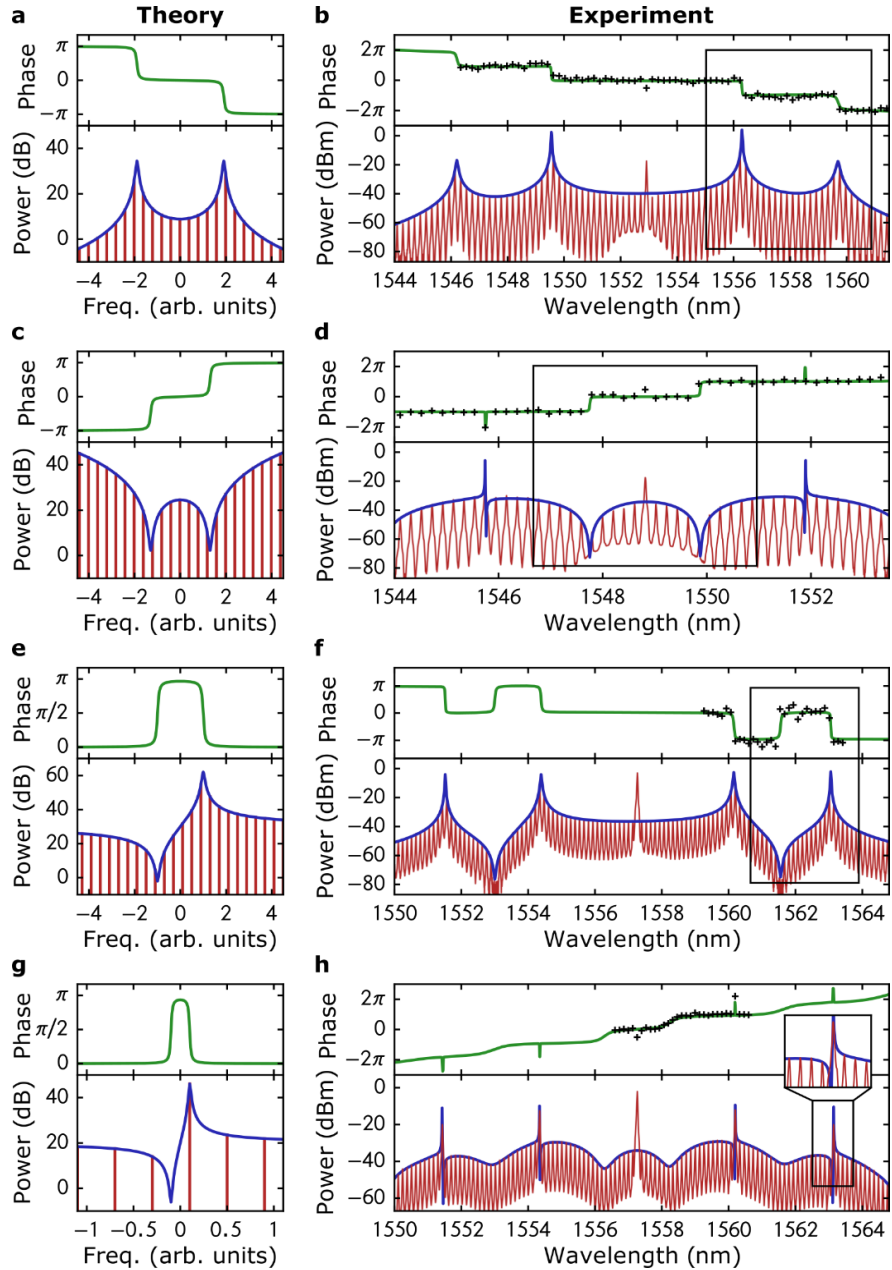

**Figure 3** Spectral features in the superoscillator model. The first column shows theoretical calculations of phases (green, upper panels) and power spectral densities (blue, envelope functions, lower panels) for different spectral features. The second column shows measurements of actual microcombs (red, comb lines) as well as phase measurements for sections of the comb (black crosses). The data is fitted and in excellent agreement with the superoscillator model (blue and green graphs). (**a,b**) Two positive superoscillators, or poles, (**c,d**) two negative superoscillators, or zeros, (**e,f**) one negative and one positive superoscillator, and (**g,h**) a negative and positive superoscillator separated by less than one comb mode spacing. The sections of the comb spectra containing the features in the left-hand column are highlighted with black rectangles. Combs (b) and (d) were generated in a silica microrod resonator and (f) and (h) in a silica disk resonator. The similarity of (d) and (h) indicates that this model applies equally well to both resonator geometries. The pump mode is the same in (f) and (h), although the coupling conditions (pump power and detuning, taper position) are slightly different. The closest superoscillator to the pump is one third of a free spectral range closer in (f) than in (h), which is due to the different pump detunings leading to vertically shifted detuning parabolas.

Equation (3) is generated by a single anomalous spacing of $(1 + \alpha)\tau$, whilst equation (4) comes from two neighboring anomalous spacings of $(1 + \alpha + \beta)\tau/2$ and $(1 + \alpha - \beta)\tau/2$. It has been suggested that the solitons' spacings could be stabilized by locking onto a "grid" formed by beating between the pump and an anomalously powerful sideband[16], such that the regular soliton spacing $\tau$ is an integer multiple $n$ of the grid spacing. If this is the case, we would expect $f_{PS}/f_{rep}$ and either $(1 + \alpha)$ or both $(1 + \alpha + \beta)/2$ and $(1 + \alpha - \beta)/2$ all to be multiples of $1/n$. We find this indeed to be true, with $n = 3$ for Fig. 3b, f and h and $n = 2$ for Fig. 3d. The mechanism for solitons locking onto the grid could be linked to soliton plasticity, which has recently been demonstrated in fiber loop resonators [25] and even used to store binary data[26]. This time-domain approach is complemented by the frequency-domain model, which may provide an alternative physical basis for superoscillators.

| Panel of Fig. 3 | (b) | (d) | (f) | (h) |
|---|---|---|---|---|
| Equation fitted | (3) | (4) | (3) | (4) |
| $f_{rep}$ (GHz) | 25.66 | 25.67 | 16.45 | 16.45 |
| $f_{PS}$ (GHz) | 419.1 | 385.5 | 356.3 | 361.6 |
| $f_{PS}/f_{rep}$ | 16.33 | 15.02 | 21.66 | 21.98 |
| $\alpha$ | 0.339 | 2.021 | 0.675 | 2.011 |
| $\beta$ | N/A | -0.06 | N/A | 0.35 |

**Table 1** Fit parameters $f_{PS} = 1/\tau$, $\alpha$ and $\beta$ of the response functions in (3, 4) to the comb amplitude spectra shown in Fig. 3, as well as the comb repetition rate $f_{rep}$. $f_{PS}$ is the frequency difference between consecutive positive superoscillators. Estimated uncertainties are of the order of the least significant quoted digit. Observe how $f_{PS}/f_{rep}$, $\alpha$ and $\beta$ are all close to an integers or simple fractions.

## IV. CONCLUSIONS

In conclusion, we introduce the concept of superoscillators that describe microresonator frequency comb generation in the frequency domain. The superoscillators are collective excitations across a large number of frequency comb modes that behave like single harmonic oscillators. We present experimental measurements of microcomb states that are in excellent agreement with the theory. The model of superoscillators in microcombs might help to reduce the complexity of microcomb states in the frequency domain. Our results represent an empirical frequency-domain description for phase-locked microcombs for both phase and amplitude. In addition, these findings generate a link between well-established theories of frequency response functions and photonic systems. Applying concepts of transfer functions [19] that are well established in the fields of signal processing, electronic networks, mechanical systems and communication theory could enable a better understanding of complex optical processes.

## APPENDIX 1: DERIVATION OF EQUATIONS (3) AND (4)

Many of the phase-locked comb states have recently been shown to correspond in the time domain to regular crystals of solitons with one or two defects such as a shifted soliton or an anomalous spacing between neighboring solitons [16]. Since such a pattern is the convolution of a single soliton with a series of Dirac delta functions at the positions of the centers of the solitons in the crystal, its spectrum will be equal to the hyperbolic-secant spectrum of a single soliton multiplied by the Fourier transform of the series of delta functions. It is this Fourier transform that will form the basis of equations (3) and (4).

Since all phase-locked comb states in finite-sized resonators have discrete spectra with a frequency spacing equal to the repetition rate of the circulating pattern, some tweaks are necessary in order to produce continuous spectra. This can be achieved by increasing the resonator size while simultaneously adding solitons to keep their spacing constant. At the same time, the defects in the soliton lattice are being kept constant. This leads to a transition into a continuous envelope of the comb spectrum while the comb mode spacing decreases, as shown in Figure 4. By this extension, the envelope itself is given by the spectrum of the same group of defects in an infinite lattice of the same spacing.

The lattices corresponding to equations (3) and (4) are illustrated in Figure 5. $H(\omega)$ is the inverse Fourier transform of a series of delta functions at the demarcated times $t_j$, which corresponds to $\sum_j \exp(i\omega t_j)$. For equation (3) this equals

$$\sum_{n=0}^{\infty} \exp(i\tau\omega(n + (1+\alpha)/2)) + \text{c.c.} = -\frac{\sin(\alpha\tau\omega/2)}{\sin(\tau\omega/2)} \qquad (6)$$

while in equation (4) $\exp(i\beta\tau\omega/2)$ is added to this (from the additional delta function).

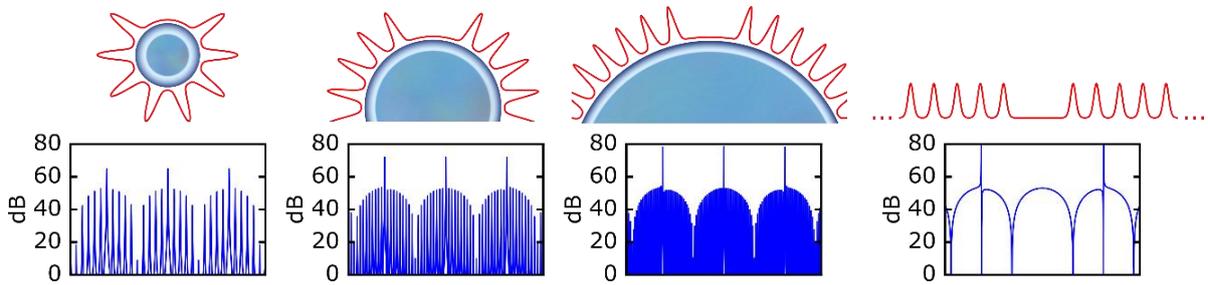

**Figure 4** Derivation of the continuous envelope of a comb spectrum. For a comb resulting from a localized defect or group of defects in an otherwise regular lattice of solitons, increasing the size of the resonator whilst adding more solitons to the regular lattice so as to keep the lattice spacing constant means decreasing the comb spacing whilst keeping the same envelope. Taking this to its natural conclusion, we find that the envelope itself corresponds to the same defect in an infinite regular lattice of the same spacing. The defect shown here is an anomalous spacing of 3.01 times the lattice spacing. From left to right: 8, 18, 38 and an infinite number of solitons. The hyperbolic secant factor corresponding to the spectrum of a single soliton has been omitted from the spectra, meaning that these are in fact Fourier transforms of series of Dirac delta functions at the centers of the solitons, and that the envelope is what we have defined as the response function $H(\omega)$. The form of $H(\omega)$ here is that of equation (3), where $\tau$ is the spacing of the regular lattice in time and $\alpha = 2.01$.

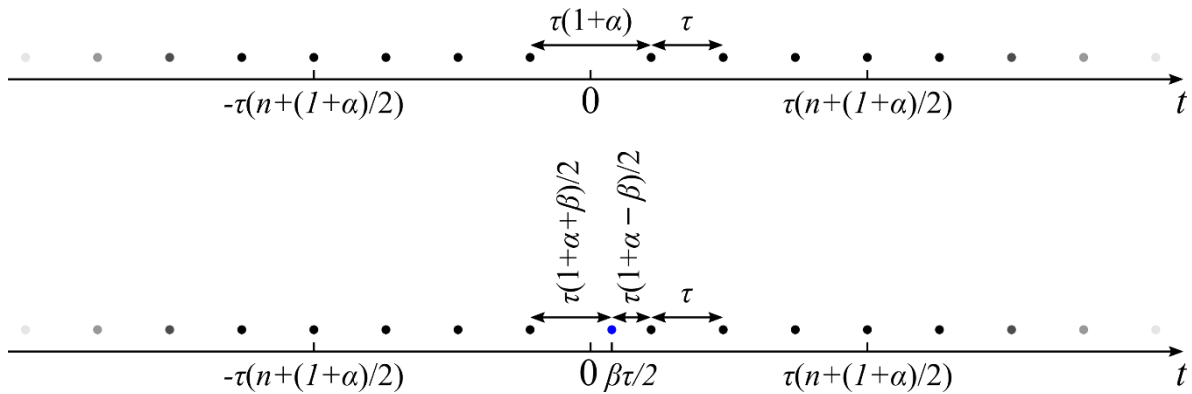

**Figure 5** Positions of Dirac delta functions in the time domain that correspond to equations (3) (top) and (4) (bottom) in the frequency domain. Both are an infinite lattice with spacing $\tau$ and an defect with a spacing of $\tau(1 + \alpha)$. The lower graph contains an additional delta function that is offset by $\beta\tau/2$ from the center of the anomalous spacing.

# APPENDIX 2: FITS OF FUNCTIONS BASED ON EQUATIONS (3) AND (4) TO COMB SPECTRA IN FIG. 3

Functions of the following form, where $H(\omega)$ is modelled corresponding to equations (3) or (4), were fitted on a dB scale to the comb amplitude spectra shown in Figure 3:

$$R(\omega) = H(\omega - \omega_\text{P}) \operatorname{sech}\left(\frac{\omega - \omega_\text{S}}{\Delta\omega_\text{S}}\right) \quad (7)$$

The fitting parameters were $\omega_\text{P}$, $\omega_\text{S}$ and $\Delta\omega_\text{S}$ as well as $\tau$, $\alpha$, $\beta$ and $K$ from equations (3) and (4). The fits are shown in Figure 6, as well as the measured phases and those calculated from the fits.

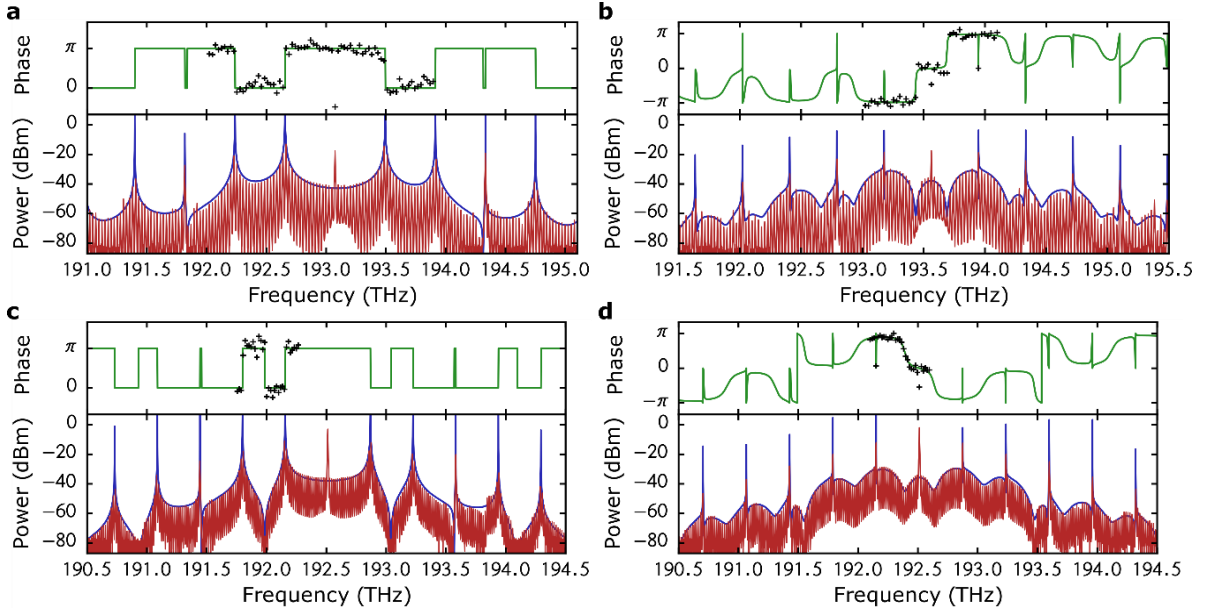

**Figure 6** Fits (blue, envelope functions, lower panels) of equation (7), i.e. equations (3) or (4) with an arbitrary frequency offset, multiplied by a hyperbolic secant with arbitrary width and frequency offset, to the peaks of the comb power spectra (red, comb lines, lower panels) shown in Figure 3. Also shown are the phase spectra (green, upper panels) calculated from the fits and some measured comb line phases (black crosses). The fits are performed on a frequency rather than wavelength scale to ensure the perfect regularity of spacings necessary for a good fit over the entire spectrum. The combs are presented in the same order as in Figure 3.

## ACKNOWLEDGEMENTS

This work has been supported by the National Physical Laboratory Strategic Research Programme. LDB is supported by the Engineering and Physical Sciences Research Council (EPSRC) through the Centre for Doctoral Training in Applied Photonics. The data in Figure 3b,d,f,h were measured at the National Institute of Standards and Technology, Boulder, USA and we thank Scott Diddams for early contributions to this work and for giving us permission to use this data.